\documentclass[aps,prl,reprint,superscriptaddress,nofootinbib,longbibliography]{revtex4-2}
\usepackage[T1]{fontenc}
\usepackage{amsmath,amssymb,bm,mathtools}
\usepackage[colorlinks=true,linkcolor=blue,citecolor=blue,urlcolor=blue]{hyperref}
\hypersetup{pdftitle={Gravitational Constraints on the SU(2) Spectrum},pdfauthor={},pdfsubject={Gravitational dispersion relations, SU(2) representations, and crossing constraints}}

\newcommand{\Cat}{\mathrm{Cat}}
\newcommand{\Id}{\mathbf 1}
\newcommand{\ee}{\mathbf e_0}
\newcommand{\cS}{\mathcal S}
\newcommand{\M}{\mathsf M} 
\newcommand{\Tr}{\operatorname{Tr}}

\newcommand{\Span}{\operatorname{span}}
\newcommand{\beq}{\begin{equation}}
\newcommand{\eeq}{\end{equation}}
\begin{document}
\title{Half-Completeness from Standard Gravitational Dispersion Relations}

\author{Lukas W. Lindwasser}
\email{llindwasser@ntu.edu.tw}
\affiliation{Department of Physics and Center for Theoretical Physics,
National Taiwan University, Taipei 10617, Taiwan}
\author{Laurentiu Rodina}
\email{laurentiu.rodina@gmail.com}
\affiliation{Beijing Institute of Mathematical Sciences and
Applications (BIMSA), Beijing 101408, China}

\author{Justinas Rumbutis}
\email{jr3918@as.edu.tw}
\affiliation{Department of Physics and Center for Theoretical Physics,
National Taiwan University, Taipei 10617, Taiwan}
\affiliation{Institute for Mathematics, Academia Sinica, Taiwan}

\begin{abstract}
The completeness hypothesis states that quantum gravity realizes
every representation allowed by its gauge group. Previous S-matrix arguments used the graviton pole to constrain exchanged
representations, including a proof of completeness by imposing an
additional high-energy bound on crossing-odd amplitudes.
We investigate what follows without this assumption, using
standard twice-subtracted dispersion relations, positivity,
and crossing symmetry. For external massless SU(2) isospin $J$, we prove
``half-completeness'': either massive exchanges include every
isospin $j\le J$, or at least one $j>J$ is required.
At $J=1$, we construct an explicit amplitude with only $j=0,1$
exchanges that satisfies these conditions but violates the 
crossing-odd bound.
This establishes that the additional high-energy assumption
does not follow from the standard framework.
For $J\ge2$, however, we show that lower-only spectra cannot
be realized within a class of string-inspired Euler amplitudes,
leaving their existence an open question.
More generally, for scalar matter transforming under any
compact internal symmetry group, we prove that massive
singlets of unbounded Lorentz spin must be exchanged.
\end{abstract}

\maketitle

\section*{Introduction}

Swampland conjectures seek to distinguish low-energy theories that admit a consistent quantum-gravitational completion from those that do not~\cite{VafaSwampland,PaltiSwampland}. In this work we focus on the completeness hypothesis, which requires every representation allowed by the gauge group to be realized by physical states~\cite{Polchinski,BanksSeiberg,HarlowOoguri,Noninvertible}. We investigate how much of this spectrum follows from scattering consistency of the S-matrix bootstrap framework~\cite{Adams2006,EFTHedron,Tolley,Extremal,IRrefStoryO,IRrefEnergyPositivity,IRrefPositiveMoments,IRrefBridgingBootstrap,IRrefScalarIsland,IRrefQuantumGravity,IRrefCurvatureSquared,IRrefEinsteinGravity,IRrefGravitonWaves,IRrefGravityIslands,IRrefSpinningPositivity,IRrefConvexGeometry,IRrefCausalitySymmetry,IRrefMultifieldPositivity,IRrefHigherPoint,IRrefLongRange,IRrefFinitePlanck,IRrefNonForward,IRrefCrossingGravity,IRrefReggeBounds,Sharp,Adding,UVrefMultipositivity,UVrefReggeBootstrap,IRrefGravitonCoupling,IntoEFTHedron,SpectrumBounds,ReverseBootstrap,Pole,UVrefNumberReggeTrajectories,UVrefGravitonReggeTrajectories,UVrefLinearTrajectories,Rodina:2026eji}. This approach has been previously used to test swampland conjectures~\cite{CheungRemmenWGC,HamadaWGC,IRrefPositivityWGC,CausalityWGC,BoundingWGC,IRrefSuperGravity,Tokareva:2025rta,IRrefUniversalCutoff,Huang:2026ute} and even to reconstruct string amplitudes and constrain their deformations~\cite{CarvingStrings,BootstrapStrings,AlmostNothing,IRStringConsistency,UVrefFlattening,UVrefCornersIslands,UVrefTripleProduct,Geiser:2022exp,UVrefVenezianoVariations,UVrefStringyDynamics,UVrefBespokeDualResonance,UVrefStringyBootstrap,UVrefTreeStringUnitarity,UVrefBespokeUnitarity,UVrefWhereString,UVrefWhereTreeString,UVrefExtremalScalars,UVrefSplittingIslands,UVrefMaximalSupersymmetry,UVrefVirasoroShapiro,UVrefMultiparticleFactorization,UVrefHigherSpinHigherPoint,UVrefStringyCompletions,UVrefMassiveHigherSpins,Chiang:2026ubv}. 

Gravity couples universally to energy and momentum, so any two particles can exchange a graviton. In elastic scattering of massless scalars, this gives a contribution proportional to $G_Ns^2/t$, where $G_N$ is Newton's constant, $s$ is the squared center-of-mass energy, and $t$ is the squared momentum transfer. The pole at $t=0$ obstructs positivity bounds in the forward limit, requiring different approaches such as smearing~\cite{Sharp} and sampling~\cite{Sampling}.

The graviton pole can also be used to anchor a different type of approach relating IR consistency to constraints on the UV spectrum. Through the twice-subtracted dispersion relation, massive exchanges must reproduce the graviton pole. When considering the scattering of colored particles, this leads to powerful conditions of the irreps that must be present as UV states, even without imposing positivity or crossing null constraints~\cite{Hillman2025}. Later work established completeness of the Cartan charge lattice
for several nonabelian groups, leaving open whether every allowed
irreducible representation must also occur; the argument remained
inconclusive for SO(3)~\cite{Calisto2025}.
 Subsequent work required every representation
generated by tensor products of the seed states, assuming improved
high-energy behavior of crossing-odd amplitudes uniformly near
the forward limit~\cite{Calisto2026}.

In this work we ask the natural question: can completeness be proven using only the standard assumptions of the S-matrix bootstrap? This includes twice-subtracted dispersion relations, positivity, and crossing symmetry, without imposing improved high-energy behavior or any constraints on the spectrum. Our goal is to understand this question in the simplest non-abelian group, and if the answer is negative, perhaps understand more generally what extra constraints are required.

For tree-level scattering of massless SU(2) scalars of isospin
$J$ we prove that
\begin{equation}
 \boxed{
 \begin{gathered}
 \text{all lower isospins }j=0,1,\ldots,\lfloor J\rfloor\text{ are exchanged}\\
 \text{or}\\
 \text{at least one higher isospin }j>J\text{ is exchanged}.
 \end{gathered}
 }
 \label{eq:alternative}
\end{equation}
Our all-$J$ analysis of the graviton-pole argument~\cite{Hillman2025}
leaves four exceptional omissions for integer $J$, all excluded
by positivity and crossing null constraints. For half-integer $J$,
the graviton relation alone suffices.

As in~\cite{Calisto2025}, one could next consider the scattering of SU(2) representations that are present as exchanged states in the massless all isospin $J$ amplitude discussed above, deriving new constraints on the spectrum consistent with the graviton-pole argument, positivity and crossing null constraints. Full completeness of the SU(2) representation spectrum would follow recursively if at least one higher isospin $j>J$ in addition to the lower isospins was necessary for consistency. Instead we find that standard assumptions are not enough to establish this.

Whether the lower-isospin branch admits an explicit amplitude is a
further question. At $J=1$, we construct a crossing-symmetric amplitude with positive
partial-wave residues and massive exchanges only in $j=0,1$. This example also shows that the improved high-energy
assumption used in Ref.~\cite{Calisto2026} does not follow
from the four-point consistency conditions imposed here. 

At higher $J$, our finite numerical tests find no obstruction to
lower-only exchange. However, we show that within a class of string-inspired
amplitudes, matching the singlet graviton pole requires an
exchanged isospin above $J$ for every $J\ge2$.

Whether $j\le J$ amplitudes exist outside this class, or are
excluded by the full dispersive constraints, remains open.
Resolving this question would clarify how much of completeness
follows from scattering consistency alone. Understanding which
properties cause the string-inspired obstruction may help identify
what other additional assumptions are needed to pass from half-completeness
to full completeness.

Independently, we show that the same framework requires massive
singlets of unbounded Lorentz spin for any compact
internal symmetry group. The unbounded spin required by gravity
therefore cannot be carried entirely by charged exchanges.
A neutral tower must couple to the scalar pair, regardless of
which charged representations are present, fixing a universal
feature of the ultraviolet spectrum.










\section*{Graviton pole and color projectors}
We consider identical massless Lorentz scalars of integer isospin $J$ in $D\geq4$ dimensions, with unbroken SU(2) symmetry and a nonconfining vacuum when it is gauged. We assume a tree-level ultraviolet completion with isolated massive poles above a gap $M_n\geq\Delta>0$. We use positive pole residues, crossing symmetry, and the standard twice-subtracted dispersion framework, including its higher-moment crossing sum rules~\cite{Sharp}.  The external pair decomposes as $J\otimes J=\bigoplus_{j=0}^{2J}j$, where $j$ is the exchanged isospin of the UV state. We denote Lorentz spin as $\ell$, so Bose symmetry requires $j+\ell$ even. The Mandelstam variables obey $s+t+u=0$, and we take $t=0$ as the forward limit. 

The amplitude carries incoming SU(2) indices $a,b$ and outgoing indices $c,d$, called color indices. The $s$ channel pairs $a$ with $b$, the $t$ channel pairs $a$ with $d$, and the $u$ channel pairs $a$ with $c$. The partial-wave expansion now involves both Lorentz spin and internal isospin~\cite{Hillman2025}:
\begin{equation}
 \mathcal A^{abcd}(s,t)=
 \sum_{j,\ell}a_{j,\ell}(s)(\mathbb{P}_j^s)^{abcd}
 G_\ell^{(D)}(1+2t/s).
 \label{eq:main-color-partial-waves}
\end{equation}
Here $G_\ell^{(D)}$ is the Gegenbauer polynomial normalized by $G_\ell^{(D)}(1)=1$, and $a_{j,\ell}$ are the partial-wave coefficients, with positive normalization factors absorbed into them. The projector $\mathbb{P}_j^s$ selects the two-particle states with total isospin $j$. Thus a massive pole in $a_{j,\ell}$ describes an exchanged particle of isospin $j$ and Lorentz spin $\ell$. As operators on the pair space, the projectors obey $\mathbb{P}_j\mathbb{P}_k=\delta_{jk}\mathbb{P}_j$ and $\sum_j\mathbb{P}_j=\Id$. We call the states of a given $j$ an isospin sector. Their pole weights $r_{n,j,\ell}$ are defined by $a_{j,\ell}(s)\sim r_{n,j,\ell}/(M_n^2-s)$ and are nonnegative when positivity is imposed.

For an explicit illustration, take external isospin $J=1$ in a real Cartesian basis, so $a,b,c,d=1,2,3$. The singlet, triplet, and quintet projectors are
\begin{align}
 (\mathbb{P}_0^s)^{abcd}&=\frac13\delta_{ab}\delta_{cd},\nonumber\\
 (\mathbb{P}_1^s)^{abcd}&=\frac12(\delta_{ad}\delta_{bc}-\delta_{ac}\delta_{bd}),\nonumber\\
 (\mathbb{P}_2^s)^{abcd}&=\frac12(\delta_{ad}\delta_{bc}+\delta_{ac}\delta_{bd})
                  -\frac13\delta_{ab}\delta_{cd}.
 \label{eq:main-delta-projectors}
\end{align}
The outgoing pair is ordered as $(d,c)$, so its identity tensor is $\delta_{ad}\delta_{bc}$. The three projectors extract the trace, antisymmetric, and symmetric-traceless parts of that pair.

Changing the exchanged channel changes which indices are paired. In this convention,
$(\mathbb{P}_j^u)^{abcd}=(\mathbb{P}_j^s)^{acbd}$ and
$(\mathbb{P}_j^t)^{abcd}=(\mathbb{P}_j^s)^{adbc}$.
Expressing the $s$- and $u$-channel tensors in the $t$-channel basis lets us compare their contributions with gravity. 

A $t$-channel graviton carries no SU(2) charge. It contributes $g_Gsu\,\mathbb{P}_0^t/t$, where $g_G>0$ includes the singlet-projector normalization. Its quadratic coefficient in $s$ is therefore $-g_G\mathbb{P}_0^t/t$. After removing the constant and linear subtraction terms, the twice-subtracted dispersion relation matches this quadratic contribution to a sum over massive poles (Appendix~\ref{sec:dispersion}). The massive exchanges must therefore reproduce a pure $t$-channel singlet.

At $s=-t/2$, we have $s=u$ so the direct and crossed channels enter symmetrically, and the massive-pole denominators coincide. Crossing gives these poles the same spectral weight, so the relation becomes
\begin{equation}
 \sum_j F_j(q)(\mathbb{P}_j^s+\mathbb{P}_j^u)
 =\frac{g_G}{q^2}\mathbb{P}_0^t+O(1),\qquad q^2=-t>0.
 \label{eq:main-projector-DR}
\end{equation}
The function $F_j(q)$ is the contribution of all massive poles of isospin $j$ to this dispersion relation:
\begin{equation}
 F_j(q)=\sum_{n,\ell}
 \frac{r_{n,j,\ell}G_\ell^{(D)}(1-2q^2/M_n^2)}
 {M_n^2(M_n^2-q^2)(M_n^2-q^2/2)}.
 \label{eq:main-Fj}
\end{equation}
The sum includes every mass and Lorentz spin in sector $j$. If isospin $j$ is absent, $F_j=0$. The $O(1)$ remainder in Eq.~\eqref{eq:main-projector-DR} stays bounded as $q\to0$. Masses and couplings change these weights but not the projector conversion coefficients. Therefore, we have the following constraint:
\noindent\fbox{%
  \parbox{\linewidth}{%
   If the allowed sums $\mathbb{P}_j^s+\mathbb{P}_j^u$ cannot reproduce $\mathbb{P}_0^t$, no regular low-energy interaction can repair the unmatched graviton pole.
  }%
}\\

Let us consider a simple example that illustrates our main result. Consider a setup with $J=1$, but where the massive exchanges have only $j=0,1$, with the higher isospin $j=2$ absent. The explicit projectors above give
\begin{align} \mathbb{P}_0^s+\mathbb{P}_0^u&=\tfrac23\mathbb{P}_0^t+\tfrac23\mathbb{P}_2^t,\nonumber\\
 \mathbb{P}_1^s+\mathbb{P}_1^u&=2\mathbb{P}_0^t-\mathbb{P}_2^t.
 \label{eq:integer-example-t-projectors}
\end{align}
 Neither remaining tensor is purely a $t$-channel singlet, so singlet exchange alone, or triplet exchange alone, cannot reproduce gravity. With both present, cancellation of $\mathbb{P}_2^t$ and matching of $\mathbb{P}_0^t$ fix
\begin{equation}
 \mathbb{P}_0^t=\tfrac12(\mathbb{P}_0^s+\mathbb{P}_0^u)
             +\tfrac13(\mathbb{P}_1^s+\mathbb{P}_1^u).
 \label{eq:integer-example-t-inverse}
\end{equation}
The solution is unique. On the other hand, this also implies that omitting either lower isospin requires an isospin-two contribution to reproduce the graviton singlet correctly.

The pole makes this stronger than a statement about isolated particles. Inverting the two component equations gives
$F_0=g_G/(2q^2)+O(1)$ and $F_1=g_G/(3q^2)+O(1)$.
Any finite set of massive poles is regular at $q=0$, so both sectors need infinitely many coupled states. Unbounded Lorentz spin will follow separately when we also impose positivity.











\section*{The all-isospin constraint}

For general integer $J$, we make the same change of projector basis:
$\mathbb{P}_j^s=\sum_K(M_{s,t})_{jK}\mathbb{P}_K^t$, where $K$ is the $t$-channel isospin.
This is SU(2) \emph{recoupling}: the Wigner $6j$ symbol gives the overlap between the two ways of pairing four SU(2) indices. The conversion coefficients are
\begin{align}
 (M_{s,t})_{jK}&=(-1)^j(2j+1)
\begin{Bmatrix}J&J&j\\J&J&K\end{Bmatrix}\nonumber\\
 &=\frac{(-1)^K(2j+1)}d R_j(y_K).
 \label{eq:main-sixj-conversion}
\end{align}
where $d=2J+1$, $y_K=K(K+1)$, and
 $R_j(y)$ is a Racah polynomial of degree $j$, normalized by $R_j(0){=}1$~\cite{DLMFProof}. For example, $R_1(y)=1-\frac{y}{2J(J+1)}$ for $J\geq1$. The crossed conversion is $(M_{u,t})_{jK}=(-1)^K(M_{s,t})_{jK}$, so $\mathbb{P}_j^s+\mathbb{P}_j^u$ has only even components, just as we saw for $J=1$ in Eq.~(\ref{eq:integer-example-t-projectors}). We can write these components explicitly as
\begin{equation}
 \begin{gathered}
 \mathbb{P}_j^s+\mathbb{P}_j^u=\sum_{k=0}^J C_{j,2k}\mathbb{P}_{2k}^t,\\
 C_{j,2k}\equiv\frac{2(2j+1)}d R_j(y_{2k}).
 \end{gathered}
 \label{eq:main-C-definition}
\end{equation}
The vector $C_j=(C_{j,0},C_{j,2},\ldots,C_{j,2J})$ simply lists the coefficients of this tensor in the $t$-channel basis. Working with these lists turns the tensor-matching problem into a finite system of linear equations.

When no higher-isospin tower is present, the singular part must be built from the $J+1$ lower tensors. Their coefficients are polynomials of successive degrees $0,\ldots,J$, evaluated at $J+1$ distinct values $y_0,y_2,\ldots,y_{2J}$. They are therefore linearly independent, giving a unique expansion
\begin{equation}
 \ee=\sum_{j=0}^{J}w_j^{(J)}C_j,
 \qquad \ee=(1,0,\ldots,0).
 \label{eq:inverse}
\end{equation}
The vector $\ee$ represents $\mathbb{P}_0^t$: its singlet coefficient is one and every nonsinglet coefficient is zero. Thus the polynomial on the left below must equal one at $y=0$ and vanish at $y_2,y_4,\ldots,y_{2J}$:
\begin{equation}
 \begin{aligned}
\frac2d\sum_{j=0}^J(2j+1)w_j^{(J)}R_j(y)=p_J(y),\\
 p_J(y)\equiv \prod_{k=1}^J\left(1-\frac{y}{2k(2k+1)}\right).
 \end{aligned}
 \label{eq:main-integer-interpolation}
\end{equation}
These $J+1$ values determine a degree-$J$ polynomial uniquely. The product supplies its zeros and normalization, and so its expansion in the Racah basis then determines every $w_j^{(J)}$.

Inverting the dispersion relation gives $F_j(q)=g_Gw_j^{(J)}/q^2+O(1)$. If $w_j^{(J)}\ne0$, omitting isospin $j$ sets $F_j=0$ and leaves an unmatched pole. In other words, an isospin sector $j$ can only be removed if its corresponding weight $w_j^{(J)}$ vanishes.  Changing the masses or couplings of the remaining lower isospins cannot fill the gap, so every nonzero weight requires its own tower.

The remaining question is whether any of the $w_j^{(J)}$ vanish in the unique expansion.  In Appendix~\ref{sec:allJ} we prove that for $J\ge 6$, all $w_j^{(J)}$ are non-vanishing. For smaller $J$, we do find precisely four cases where $w_j^{(J)}=0$:
$(J,j)=(2,0),(3,1),(3,2),(5,2)$.
These four zeros identify the only lower-isospin omissions left by this equation. Remarkably, positivity and crossing null constraints are sufficient to remove these exceptions, proving half-completeness for integer $J$.

\paragraph*{Half-integer external isospin.}
For half-integer $J$, the corresponding
graviton-matching problem again has a unique
solution, but this time with no vanishing lower-isospin
coefficients. If neither channel contains massive
exchanges with $j>J$, the graviton relation alone
therefore requires a tower of unbounded mass in
every $j=0,\ldots,\lfloor J\rfloor$ in each
channel. 

Details are given in Appendix~\ref{sec:half-integer}. The accompanying Wolfram Language code also computes these weights
for any specified integer or half-integer $J$.

\section*{Positivity and crossing}
In this section we explain how specially constructed null constraints can eliminate the four exceptions that appear for integer $J$.
 Crossing null constraints are obtained by expanding
the amplitude at low energies and
matching its Taylor coefficients before and after
exchanging external legs. Replacing these coefficients
by their dispersive expressions gives relations among
massive exchanges, which directly constrain the UV spectrum. The complete derivations are done explicitly in Appendices~\ref{sec:dispersion} and \ref{sec:singlet}, here we consider the $J=2$ case, where the above discussion allows the massive $j=0$ state to be absent.

Define the weighted pole sums, and for simplicity set $D=4$
\begin{equation}
 \begin{gathered}
 \mathcal N_j=
 \sum_{n,\ell}\frac{r_{n,j,\ell}}{M_n^{14}}
 \pi_j(\mathcal{J}_\ell),\qquad
 \mathcal{J}_\ell=\ell(\ell+1),\\
 \pi_j(x)=
 \begin{cases}
 x(x-6),&j\ \text{even},\\
 (x-2)(x-12),&j\ \text{odd}.
 \end{cases}
 \end{gathered}
 \label{eq:main-singlet-moments}
\end{equation}
The mass factor comes from expanding the pole
denominators, and the spin polynomials from expanding
their angular dependence. Crossing requires
\begin{equation}
 4\mathcal N_0
 =9\mathcal N_1+5\mathcal N_2
  +21\mathcal N_3+9\mathcal N_4.
 \label{eq:j2-degree-six}
\end{equation}

Bose symmetry allows even spins for even $j$ and
odd spins for odd $j$. On these allowed spins,
the even-isospin weight vanishes at $\ell=0,2$
and is positive above them. The odd-isospin weight
vanishes at $\ell=1,3$ and is positive above them.
Positive pole residues therefore give
$\mathcal N_j\ge0$ for every isospin.

By itself, Eq.~\eqref{eq:j2-degree-six} would allow
every moment to vanish, with all exchanged spins
at most three. The graviton pole excludes this
possibility. With positive residues, reproducing
its $1/q^2$ singularity requires arbitrarily high
Lorentz spins, even if infinitely many masses are
available. 

If the singlet were absent this would imply $\mathcal N_0=0$,
forcing every term on the
right-hand side of Eq.~\eqref{eq:j2-degree-six}
to vanish. This contradicts the spin requirement, and
so a massive singlet with
$\ell\ge4$ must be present. 

Analogous null constraints exclude the remaining
omissions when all $j>J$ exchanges are absent:
$j=1,2$ at $J=3$ and $j=2$ at $J=5$.
Their explicit forms and sign arguments are given
in Appendix~\ref{sec:nulls}. All four exceptions
are therefore removed, completing the proof of
Eq.~\eqref{eq:alternative}.

\section*{A universal singlet tower}
Gravity requires unbounded spin somewhere in the coupled spectrum. In this section we show that crossing symmetry implies something stronger. A bound on singlet spin would bound every coupled spin, even when higher isospins are present.

The $J=2$ example above required a massive singlet with $\ell\ge4$.
Continuing in $D=4$, we now show that singlets of arbitrarily
high spin must occur for every integer $J\ge1$, regardless of
which other isospins are exchanged.

To test any proposed bound on singlet spin, we need null constraints
whose weights vanish below that bound and are positive above it.
Following an argument similar to Ref.~\cite{IntoEFTHedron},
Appendix~\ref{sec:singlet} constructs such constraints for every
integer $p\ge2$. Define
\begin{equation}
 \mathcal N_j^{(p)}
 =\sum_{n,\ell}\frac{r_{n,j,\ell}}{M_n^{6p+2}}
   \prod_{a=0}^{p-1}
   [\mathcal J_\ell-\mathcal J_{\eta_j+2a}],
 \label{eq:main-singlet-p-moments}
\end{equation}
where $\mathcal J_\ell=\ell(\ell+1)$ and $\eta_j=j\bmod2$.
Crossing gives
\begin{equation}
 2J\,\mathcal N_0^{(p)}
 =\sum_{j=1}^{2J}(2j+1)(2-(-1)^j)
   \mathcal N_j^{(p)}.
 \label{eq:main-singlet-p-balance}
\end{equation}
The product removes the first $p$ allowed spins: $0,2,\ldots,2p-2$
for even isospin and $1,3,\ldots,2p-1$ for odd isospin.
Every higher allowed spin has positive weight, so
$\mathcal N_j^{(p)}\ge0$. At $p=2$ and $J=2$,
this reproduces Eq.~\eqref{eq:j2-degree-six}.

Suppose singlet spin were bounded, and choose $2p$ above that
bound. Then $\mathcal N_0^{(p)}=0$, so every nonnegative term
on the right must also vanish. All exchanged spins would
therefore lie among the zeros of the weights, with $\ell<2p$.
This contradicts the unbounded spin required by the graviton
pole, implying the singlet spin cannot be bounded.

\paragraph{General groups}The argument extends to any compact internal symmetry group $G$. Consider $n_V$ real scalar components. The pair
\[
 |\Omega\rangle=\frac1{\sqrt{n_V}}\sum_a|aa\rangle
\]
is invariant under $G$, so only singlet particles couple to it.
The remaining pair states are either symmetric and orthogonal
to $\Omega$, or antisymmetric.

Let $\mathcal I_{\Omega,p}$ sum the weighted squared couplings
to $\Omega$, and let $\mathcal I_{+,p}$ and $\mathcal I_{-,p}$
sum those to the remaining symmetric and antisymmetric pairs.
We use the mass factors and spin weights of
Eq.~\eqref{eq:main-singlet-p-moments}, with even spins for
$\Omega,+$ and odd spins for $-$. All three moments are
nonnegative and receive strictly positive contributions
from coupled particles with $\ell\ge2p$. Crossing gives
(Appendix~\ref{sec:universal-singlet})
\begin{equation}
 (n_V-1)\mathcal I_{\Omega,p}
 =\mathcal I_{+,p}+3\mathcal I_{-,p}.
 \label{eq:main-general-null}
\end{equation}

For SU(2), $n_V=2J+1$ and the $+$ and $-$ sectors collect the
even $j\ge2$ and the odd $j$, so Eq.~\eqref{eq:main-general-null}
reduces to Eq.~\eqref{eq:main-singlet-p-balance}. If singlet spin
were bounded, choosing $2p$ above that bound would give
$\mathcal I_{\Omega,p}=0$. Positivity would then
force both remaining moments to vanish, excluding high-spin
exchanges from every pair state. This contradicts the
unbounded spin required by the graviton pole. Singlets must
therefore have unbounded even spin.

Reference~\cite{Calisto2026} obtains a singlet tower using
an improved high-energy bound on crossing-odd amplitudes.
Here it follows from the twice-subtracted graviton relation,
positive residues, and higher-moment crossing sum rules,
without that additional bound.









\section*{Explicit amplitudes}
\label{sec:explicit-amplitudes}
We realize the lower-isospin branch explicitly at $J=1$
with a crossing-symmetric amplitude whose massive exchanges
contain only $j=0,1$. Surprisingly, this amplitude also
provides a counterexample to the improved crossing-odd
high-energy bound of Ref.~\cite{Calisto2026}, despite
having a universal graviton pole, nonnegative partial-wave
residues for $4\le D\le10$, and standard twice-subtracted
high-energy behavior.

Set the mass scale $M=1$ and the massless gauge coupling
to zero. Define the symmetric block
\begin{equation}
 \mathcal V(x,y)=(x^2+xy+y^2)
 \frac{\Gamma(-x)\Gamma(-y)}{\Gamma(-x-y)},
 \label{eq:explicit-beta-block}
\end{equation}
with poles at nonnegative integer values of $x$ and $y$.
In the triplet's Cartesian basis, take
$T_{st}^{abcd}=\delta^{ab}\delta^{cd}-\delta^{ac}\delta^{bd}
+\delta^{ad}\delta^{bc}$,
$T_{su}^{abcd}=T_{st}^{abdc}$, and
$T_{tu}^{abcd}=T_{st}^{acbd}$.
Crossing permutes the three terms in
\begin{equation}
\begin{split}
 \mathcal A^{abcd}(s,t)=\frac{g_G}{6}\big[&
 T_{st}^{abcd}\mathcal V(s,t)
 +T_{su}^{abcd}\mathcal V(s,u)\\
 &+T_{tu}^{abcd}\mathcal V(t,u)\big].
\end{split}
 \label{eq:explicit-J1-amplitude}
\end{equation}
The massless residues reproduce graviton exchange,
\begin{equation}
 \lim_{t\to0}(-t)\mathcal A(s,t)=g_Gs^2\mathbb{P}_0^t.
 \label{eq:explicit-J1-graviton}
\end{equation}
At a massive $t$-channel pole, only the $st$ and $tu$
blocks contribute. Since
$T_{st}=3\mathbb{P}_0^t-2\mathbb{P}_1^t$ and
$T_{tu}=3\mathbb{P}_0^t+2\mathbb{P}_1^t$,
the exchanged spectrum is $j=\{0,1\}$.
The poles lie at $M_n^2=n$, $n\ge1$, with unbounded
spin in both sectors. Using the results of Ref.~\cite{Mansfield2025}, one can verify that
the amplitude has nonnegative partial-wave residues for $4\le D\le10$.
It also satisfies the fixed-transfer high-energy bound required
for a twice-subtracted dispersion relation.

At large complex $s$ and fixed $t<0$,
$\mathcal V(s,t)\simeq\Gamma(-t)s^2(-s)^t$.
The remaining blocks grow no faster, giving
$\mathcal A=o(|s|^2)$.
To examine the stronger bound, define $z=s+t/2$ and
$\mathcal A^{(-)}(z,t)
=[\mathcal A(z,t)-\mathcal A(-z,t)]/2$.
For $z=iy$, $y\to+\infty$, and fixed $-1<t<0$,
\begin{equation}
 \mathcal A^{(-)}(iy,t)\sim
 \frac{i\pi g_G}{3\Gamma(1+t)\cos(\pi t/2)}
 y^{2+t}\mathbb{P}_1^t .
\end{equation}
The coefficient remains nonzero as $t\to0^-$.
Consequently, no positive, $t$-independent $\delta$
gives the suppression
$\mathcal A^{(-)}=o(|z|^{2-\delta})$
required by Ref.~\cite{Calisto2026}.

The planar open-string example in that reference
instead fails a strengthened even-sector bound and
contains no graviton exchange. Our construction
demonstrates the independence of the improved
crossing-odd bound within the gravitational
four-point setup itself.

\paragraph*{The obstruction at $J=2$.}
Now take external isospin $J=2$. The external indices run over
five components, and the projectors resolve
$2\otimes2=0\oplus1\oplus2\oplus3\oplus4$.
Keep the same beta-function blocks and delta tensors, replacing
the overall normalization $g_G/6$ by $g_G/10$.

The difference from $J=1$ is in the antisymmetric pair states.
For two triplets these have only isospin one, whereas for two
isospin-two particles they contain both $j=1$ and $j=3$.
Consequently, the same delta tensors now give
\begin{align}
 T_{st}&=5\mathbb{P}_0^t-2(\mathbb{P}_1^t+\mathbb{P}_3^t),\nonumber\\
 T_{tu}&=5\mathbb{P}_0^t+2(\mathbb{P}_1^t+\mathbb{P}_3^t).
 \label{eq:explicit-J2-delta-tensors}
\end{align}
Their sum still reproduces the graviton:
$\lim_{t\to0}(-t)\mathcal A=g_Gs^2\mathbb{P}_0^t$.
The problem appears at the massive poles.

At the first massive $t$-channel pole, $t=1$, we have
$u=-1-s$. The residues of
$\mathcal V(s,t)$ and $\mathcal V(t,u)$ are respectively
$(s+1)(s^2+s+1)$ and $-s(s^2+s+1)$. The remaining block,
$\mathcal V(s,u)$, has no $t$-channel pole here. Thus
\begin{equation}
\begin{split}
 \mathcal R_0^{(1)}(s)
 &:=\lim_{t\to1}(1-t)\mathcal A(s,t)\\
 &=\frac{g_G}{10}(s^2+s+1)
 \left[5\mathbb{P}_0^t-2(2s+1)(\mathbb{P}_1^t+\mathbb{P}_3^t)\right].
\end{split}
 \label{eq:explicit-J2-first-t-residue}
\end{equation}
The isospin-three coefficient does not vanish as a function
of scattering angle. Although the nonsinglet contributions
cancel at the massless pole, their different angular factors
prevent that cancellation at the massive pole. The same
ansatz therefore exchanges $\{0,1,3\}$, rather than only
$j\le2$.

The obstruction extends beyond this particular ansatz.
Appendix~\ref{app:explicit-dual} treats regular Euler integrals
with analytic coefficient functions and finite polynomial
momentum dependence, including arbitrarily large finite sums of
polynomially dressed, integer-shifted beta functions with a
common slope. Absence of forbidden poles at every mass level
forces each channel-pair block to obey the isospin cutoff in both
of its channels. The $t$-channel graviton pole receives
contributions only from the $st$ and $tu$ blocks, whose allowed
tensors for $j\le J$ span one direction each, and their sum
contains $\mathbb{P}_0^t$ only for $J=1$. Within this class a singlet
graviton therefore requires a massive exchange with $j>J$ for
every $J\ge2$. The existence of lower-only amplitudes outside this regular Euler class remains open. 








\section*{Conclusions}
The graviton pole constrains the exchanged representations independently of their masses and coupling strengths. For SU(2), the lower-isospin tensors form a basis, so there is no freedom to compensate for a missing nonzero coefficient by adjusting the other exchanges. Crossing and positivity remove the four remaining omissions. This uniquely fixed lower-isospin graviton expansion gives a concrete target for further completeness tests: a condition incompatible with it would force higher isospin, although filling the remaining gaps would require further constraints.

Independently of the above results, we have also numerically tested lower-only exchange for $J=1,\ldots,10$ in both $D=4$ and $D=5$. In every case, we found positive residue sums satisfying all independent crossing nulls through total Taylor degree twelve, together with a finite set of smeared twice-subtracted relations.  These finite tests find no obstruction to $j\le J$. The related code is included as an ancillary Wolfram Language code file.

Our explicit $J=1$ construction is a crossing-symmetric
meromorphic four-scalar amplitude whose massless poles
factorize through Einstein graviton exchange and whose
massive partial-wave residues are nonnegative. It satisfies the standard twice-subtracted
bound, yet exchanges only $j=0,1$ and is a counter-example to the uniform
crossing-odd high-energy bound of Ref.~\cite{Calisto2026}.
Thus the four-point assumptions used here imply neither
every representation in the external tensor product nor
that additional bound, which remains an independent additional assumption on the UV behavior.

For $J\ge2$,
however, we find the regular Euler constructions studied here cannot realize a $j\le J$ spectrum. It remains an open question whether more general meromorphic amplitudes can realize this lower-isospin branch, or whether the full bootstrap constraints exclude it. 

\vspace{0.3cm}

\paragraph*{Acknowledgments.} The authors would like to thank Yu-tin Huang for discussions and collaboration on related work.
The research of LL is supported by the Taiwan NSTC Grant No. 113-2811-M-002-167-MY3 and the Yushan Young Fellowship. LR is supported by the National Natural Science Foundation of China General Program No.~12475070 and the Beijing Natural Science Foundation International Scientist Project No.~IS24014. AI tools were used to assist with the nonvanishing argument in the all-$J$ proof, ancillary verification codes, and manuscript editing.

\bibliographystyle{apsrev4-2}
\nocite{VafaSwampland,PaltiSwampland,Polchinski,BanksSeiberg,HarlowOoguri,Noninvertible}
\bibliography{Gravitational_Constraints_SU2_Spectrum}

\clearpage
\onecolumngrid
\appendix
\setcounter{secnumdepth}{3}
\numberwithin{equation}{section}

\section{Dispersion relations and forward nulls}
\label{sec:dispersion}
The graviton relation and crossing nulls follow from the contour
identities below. Contours are counterclockwise, and each cutoff
includes complete mass levels in both crossed channels.

\subsection{Conventions and light poles}
For integer $J$ in a real basis, let $(Q_j)_{ab,cd}$ be the ordinary
orthogonal pair-space projector. Our four-leg ordering is
\begin{equation}
 (\mathbb{P}_j^s)^{abcd}=(Q_j)_{ab,dc},\qquad
 (\mathbb{P}_j^t)^{abcd}=(Q_j)_{ad,cb},\qquad
 (\mathbb{P}_j^u)^{abcd}=(Q_j)_{ac,db}.
 \label{eq:channel-index-dictionary}
\end{equation}
The reversal of the outgoing pair is only an index convention.
For squared mass $\mu=M_n^2\geq\Delta^2$ we use
\begin{equation}
 t=-q^2,\qquad s+t+u=0,\qquad
 z=s+\frac t2=\frac{s-u}{2},\qquad
 s=z-\frac t2,\quad u=-z-\frac t2.
 \label{eq:kinematics}
\end{equation}
The angular polynomial is $G_\ell^{(D)}(\xi)
=C_\ell^{(D-3)/2}(\xi)/C_\ell^{(D-3)/2}(1)$, and the residue convention is
\begin{equation}
 \mathcal A(s,t)\sim
 \frac{r_{n,j,\ell}}{\mu_n-s}G_\ell^{(D)}(1+2t/\mu_n)\mathbb{P}_j^s,\qquad
 d\rho_{j,\ell}(\mu)=\sum_n r_{n,j,\ell}\delta(\mu-\mu_n)d\mu.
 \label{eq:raw-residue}
\end{equation}
Positive factorization means $r_{n,j,\ell}\geq0$.
With the main-text conversion matrix $\M=M_{s,t}$,
\begin{equation}
 M_{u,t}=\M W,\qquad W_{KK}=(-1)^K,\qquad
 C=\M(1+W),\qquad \mathbf A(-z,t)=\mathbf A(z,t)W.
 \label{eq:conventions}
\end{equation}
A massive pole and its crossed partner in component $K$ are
\begin{equation}
 \M_{jK}G_\ell^{(D)}(1+2t/\mu)
 \left[\frac1{a_\mu-z}+\frac{(-1)^K}{a_\mu+z}\right]d\rho_{j,\ell},
 \qquad a_\mu=\mu+t/2.
 \label{eq:pole-pair}
\end{equation}

Subtract all known massless exchanges in a crossing-symmetric way,
so $\mathbf B=\mathbf A-\mathbf A_{\rm light}$ is analytic near $z=t=0$.
For graviton and gauge exchange,
\begin{align}
 \mathcal A_{\rm light}={}&g_G\left[
 \mathbb{P}_0^s\frac{tu}{s}+\mathbb{P}_0^t\frac{us}{t}+\mathbb{P}_0^u\frac{st}{u}\right]\nonumber\\
 &+g_{\rm YM}^2\left[
 \mathbb{P}_1^s\frac{t-u}{s}+\mathbb{P}_1^t\frac{u-s}{t}+\mathbb{P}_1^u\frac{t-s}{u}\right].
 \label{eq:full-light}
\end{align}
Representation normalizations are absorbed in the couplings.
At fixed $t\ne0$ its polynomial part is
$-g_Gz^2\ee/t-2g_{\rm YM}^2z\mathbf e_1/t+O(1)$.
Thus a contour enclosing $z=0$ and both massless energy poles
$z=\pm t/2$ gives
\begin{equation}
 \mathcal L_{m,K}(t):=
 \oint_{\gamma_{\rm light}}\frac{d\zeta}{2\pi i}
 \frac{A_{{\rm light},K}(\zeta,t)}{\zeta^{m+1}}
 =-\frac{g_G}{t}\delta_{m2}\delta_{K0}
  -\frac{2g_{\rm YM}^2}{t}\delta_{m1}\delta_{K1},
 \qquad m\geq1.
 \label{eq:light-moments}
\end{equation}
This includes all three light-exchange channels, not just the
Taylor series at $z=0$. Additional known light poles must be treated
on the same contour. Energy-independent scalar exchange gives no
source for $m\geq1$.

\subsection{The graviton relation}
For $0<q^2<\Delta^2/4$, put $c=q^2/2$ and define
\begin{equation}
 H_{\rm g}(\zeta,t)=\frac1{\zeta(\zeta^2-c^2)},\qquad
 \mathfrak D_t f=\frac{f(c,t)+f(-c,t)-2f(0,t)}{2c^2},\qquad
 \mathcal B^{(2)}_\Lambda=\oint_{\Gamma_\Lambda}\frac{d\zeta}{2\pi i}
 H_{\rm g}\mathbf A.
 \label{eq:Hg}
\end{equation}
The inner contour equals $-g_G\ee/t+\mathfrak D_t\mathbf B$,
because $\mathfrak D_t[z^2]=1$ and the gauge term vanishes.
At a massive pole,
$H_{\rm g}(a_\mu,t)=1/[\mu(\mu+t)(\mu+t/2)]$.
Its odd parity compensates the opposite residue signs of the
two poles in Eq.~\eqref{eq:pole-pair}, giving
\begin{equation}
 -\frac{g_G}{t}\ee+\mathfrak D_t\mathbf B
 =\sum_{j,\ell}\int_{\Delta^2}^{\Lambda^2}
 \frac{G_\ell^{(D)}(1+2t/\mu)}{\mu(\mu+t)(\mu+t/2)}
 C_j\,d\rho_{j,\ell}+\mathcal B^{(2)}_\Lambda(t).
 \label{eq:finite}
\end{equation}
The standard high-energy condition $\mathcal A(s,t)=o(|s|^2)$ on
the large complex-energy contours removes the outer contour at fixed transfer,
\begin{equation}
 \lim_{\Lambda\to\infty}\mathcal B^{(2)}_\Lambda(t)=0,
 \qquad t<0\ \hbox{fixed and sufficiently small}.
 \label{eq:fixed-boundary}
\end{equation}
With
\begin{equation}
 K_2(\mu,\ell;q)=
 \frac{G_\ell^{(D)}(1-2q^2/\mu)}
 {\mu(\mu-q^2)(\mu-q^2/2)},
 \label{eq:K2}
\end{equation}
the resulting relation is
\begin{equation}
 \frac{g_G}{q^2}\ee+\mathbf E(q)=\sum_{j=0}^{2J}F_j(q)C_j,\qquad
 F_j(q)=\sum_\ell\int_{\Delta^2}^{\infty}K_2\,d\rho_{j,\ell},
 \qquad \mathbf E(q)=\mathfrak D_{-q^2}\mathbf B.
 \label{eq:DR}
\end{equation}
The vector $\mathbf E$ and any finite massive-pole sum are regular
at $q=0$. If $\cS_\infty$ denotes the isospins with unbounded coupled
masses, annihilating their rows must also annihilate the graviton:
\begin{equation}
 \ee\in\Span\{C_j:j\in\cS_\infty\}.
 \label{eq:span}
\end{equation}
This high-energy condition is imposed on the tree amplitude itself,
not inferred from a bound on the full quantum amplitude~\cite{Pole}.

\subsection{Central moments and their null constraints}
Define
\begin{equation}
 b_{m,K}(t)=\frac{\partial_z^mB_K(0,t)}{m!}
 =\sum_{n\geq0}b_{m,n;K}t^n,\qquad
 \beta^\Lambda_{m,K}(t)=\oint_{\Gamma_\Lambda}\frac{d\zeta}{2\pi i}
 \frac{A_K(\zeta,t)}{\zeta^{m+1}}.
 \label{eq:central-def}
\end{equation}
The same contour deformation gives
\begin{equation}
 b_{m,K}(t)+\mathcal L_{m,K}(t)
 =\sum_{j,\ell}\int_{\Delta^2}^{\Lambda^2}
 \frac{\M_{jK}[1+(-1)^{m+K}]G_\ell^{(D)}(1+2t/\mu)}
 {(\mu+t/2)^{m+1}}\,d\rho_{j,\ell}
 +\beta^\Lambda_{m,K}(t).
 \label{eq:central-finite}
\end{equation}
Its denominator differs from Eq.~\eqref{eq:finite}.
For the source-free moments used here, $m\geq4$ and $m=K\bmod2$.
Writing $x=x_\ell=\ell(\ell+D-3)$ and $(a)_r$ for a rising factorial,
\begin{align}
 G_\ell^{(D)}(1+2v)&=\sum_{r\geq0}A_r(x)v^r,\qquad
 A_r(x)=\frac{\prod_{a=0}^{r-1}[x-a(a+D-3)]}
 {r!((D-2)/2)_r},\label{eq:angular-series}\\
 b_{m,n;K}&=\sum_{j,\ell}\int_{\Delta^2}^{\Lambda^2}
 \frac{2\M_{jK}Q_{mn}(x)}{\mu^{m+n+1}}d\rho_{j,\ell}
 +\epsilon^\Lambda_{m,n;K},\label{eq:finite-jets}\\
 Q_{mn}(x)&=\sum_{r=0}^n A_r(x)\left(-\frac12\right)^{n-r}
 \frac{(m+1)_{n-r}}{(n-r)!},\qquad
 \epsilon^\Lambda_{m,n;K}=\frac{\partial_t^n\beta^\Lambda_{m,K}(0)}{n!}.
 \label{eq:jetkernel}
\end{align}
These are finite-cutoff identities. A degree-$k$ crossing relation
$\sum_{m,K}v_{mK}b_{m,k-m;K}=0$ therefore gives
\begin{align}
 0&=\sum_{j,\ell}\int_{\Delta^2}^{\Lambda^2}
 \frac{\mathcal K^{(v)}_{j,\ell}}{\mu^{k+1}}d\rho_{j,\ell}
 +\epsilon_v^\Lambda,\label{eq:null-finite}\\
 \mathcal K^{(v)}_{j,\ell}
 &=2\sum_{m,K}v_{mK}\M_{jK}Q_{m,k-m}(x_\ell),\qquad
 \epsilon_v^\Lambda=\sum_{m,K}v_{mK}\epsilon^\Lambda_{m,k-m;K}.
 \label{eq:null-kernel}
\end{align}












\section{Nonvanishing of the lower-isospin coefficients}
\label{sec:allJ}

The interpolation formula \eqref{eq:main-integer-interpolation}
fixes the lower-isospin weights $w_j^{(J)}$ uniquely. This
appendix proves that none of them vanishes for integer $J\ge6$,
and for every half-integer $J$.

The proof strategy is as follows. Crossing symmetry packages all the weights into one polynomial
$h$ of degree $J$: the weight $w_j$ is read off from the value of
$h$ at the point $y_j=j(j+1)$, and $w_j=0$ corresponds to
$h(y_j)=\pm1$. The vanishing of the weights with $j>J$ tells us
the values of $h$ at the $J$ points $y_{J+1},\dots,y_{2J}$; with
the leading coefficient this determines $h$. The task is then to
show that $|h|$ exceeds one at the $J+1$ points $y_0,\dots,y_J$,
all of which lie to the left of the points where $h$ is known.
The key step rewrites $h$ as a sum of simple poles with positive
residues. For such a function the position of every root and the
sign on every interval are immediate, and two elementary
estimates reduce the whole question to three numbers, all larger
than one for $J\ge6$. The half-integer case is simpler and
needs no estimate.

\subsection{Integer isospin}
\label{sec:uniform-six}

Throughout, $d=2J+1$, $y_j=j(j+1)$, and $w_j=w_j^{(J)}$ with
$w_j=0$ for $j>J$. The points $y_j$ with $0\le j\le J$ are the
lower nodes; the points $a_r=y_r=r(r+1)$ with $J<r\le2J$ are
the upper nodes. $\Cat_J=\binom{2J}{J}/(J+1)$ is the Catalan
number.

\paragraph*{Step 1: one polynomial for all weights.}
Applying crossing to the interpolation formula
gives~\cite{DLMFProof}
\begin{equation}
 dw_j=1-(-1)^jh(y_j),\qquad
 h(y)=\sum_{k=0}^J(2k+1)(-1)^kw_kR_k(y),
 \label{eq:integer-h}
\end{equation}
with $h$ a polynomial of degree $J$. Hence $w_j=0$ if and only
if $h(y_j)=(-1)^j$, and it suffices to prove
\begin{equation}
 |h(y_j)|>1,\qquad 0\le j\le J.
 \label{eq:goal}
\end{equation}

\paragraph*{Step 2: what is known about $h$.}
At the upper nodes the weights vanish, so Eq.~\eqref{eq:integer-h}
gives $h(a_r)=(-1)^r$ for $J<r\le2J$; the leading term of the
interpolation formula gives the coefficient of $y^J$ in $h$,
\begin{equation}
 h(a_r)=(-1)^r\quad(J<r\le2J),\qquad
 [y^J]\,h=\frac{d}{2\,(2J+1)!}.
 \label{eq:hdata}
\end{equation}
These $J+1$ conditions determine $h$; the lower weights are
never computed.

\paragraph*{Step 3: positive-residue form.}
The data of Step~2 describe $h$ at points where it is not
needed. To transport this information to the lower nodes, divide
by the polynomial $Q(y)=\prod_{r=J+1}^{2J}(1-y/a_r)$ whose zeros
are exactly the upper nodes. The ratio of two degree-$J$
polynomials is a constant plus simple poles, and the known values
$h(a_r)$ become the residues:
\begin{equation}
 \mathcal R(y):=\frac{2(-1)^J\,h(y)}{\Cat_J\,Q(y)}
 =d-\sum_{r=J+1}^{2J}\frac{v_r}{a_r-y},\qquad v_r>0.
 \label{eq:partial}
\end{equation}
This is the partial-fraction (barycentric) form of Lagrange
interpolation; the constant $d$ accounts for the leading
coefficient in Eq.~\eqref{eq:hdata}. The residues are positive because the
values $h(a_r)=(-1)^r$ and the derivatives $Q'(a_r)$ alternate
in the same way, so their ratio has a fixed sign; explicitly,
$v_r=(2a_r/\Cat_J)\prod_{k\ne r}a_k/|a_k-a_r|$.

Positivity of the residues is the fact that drives the rest of
the proof. A sum of simple poles with positive residues decreases
strictly on every interval free of poles: on $(-\infty,a_{J+1})$
from $d$ to $-\infty$, and on each gap $(a_r,a_{r+1})$ from
$+\infty$ to $-\infty$. Consequently $h$ has exactly one root
$\rho_1$ below the upper nodes and one root in each of the $J-1$
gaps between them. These are all $J$ roots of $h$: they are real
and ordered as $\rho_1<a_{J+1}<\rho_2<\dots<\rho_J$.

\paragraph*{Step 4: the root $\rho_1$ lies strictly between $J$ and $J+1$.}
The numbers $p_r=v_r/(2Ja_r)$ form a
probability distribution on the upper nodes with
\begin{equation}
 \sum_rp_r=1,\qquad
 \sum_rp_ra_r=\alpha:=(J+1)(2J+1),\qquad
 \sum_rp_r(a_r-\alpha)^2=(J-1)\alpha,
 \label{eq:moments}
\end{equation}
so that $\mathcal R(y)=1-2Jy\sum_rp_r/(a_r-y)$. If all the mass
$p_r$ sat at the mean $\alpha$, this would vanish exactly at
$y=J+1$. The spread of the distribution shifts the root to the
left, and elementary estimates using the mean and the variance
show that it does not reach $J$:
\begin{equation}
 \mathcal R(J)>\frac2{(J+1)^2},\qquad
 \mathcal R(J+1)\le-\frac{(J-1)(2J+1)}{2J(4J^2+J-1)}<0.
 \label{eq:rootgap}
\end{equation}
Hence $J<\rho_1<J+1$. The open interval $(J,J+1)$ contains no
integer and therefore no node, so $h$ is nonzero at every lower
node. Equation~\eqref{eq:goal} requires the stronger bound
$|h(y_j)|>1$.

\paragraph*{Step 5: three values control all nodes.}
Since all roots of $h$ are real, $|h|$ decreases strictly on
$(-\infty,\rho_1)$, and by Rolle's theorem it rises and then
falls between consecutive roots, so its minimum over any closed
subinterval of $(\rho_1,\rho_2)$ is attained at an endpoint. The
lower nodes below $\rho_1$ satisfy $y_j\le J$; those above satisfy
$J+1\le y_j\le y_J<a_{J+1}<\rho_2$. Therefore
\begin{equation}
 \min_{0\le j\le J}|h(y_j)|\ \ge\
 \min\{\,|h(J)|,\ |h(J+1)|,\ |h(y_J)|\,\}.
 \label{eq:three}
\end{equation}

\paragraph*{Step 6: the three values.}
Since $|h|=\tfrac12\Cat_J\,Q\,|\mathcal R|$, the first two
values follow from Eq.~\eqref{eq:rootgap} and the elementary
bound $Q(y)\ge(J+1)/(2J+1)$ for $0\le y\le J+1$:
\begin{equation}
 |h(J)|\ge B_1:=\frac{\Cat_J}{(J+1)(2J+1)},\qquad
 |h(J+1)|\ge B_2:=\frac{\Cat_J\,(J^2-1)}{4J(4J^2+J-1)}.
\end{equation}
The third needs no estimate: the leading coefficient of the
interpolation formula gives
$w_J=J!\,(3J+1)!/\bigl(2[(2J+1)!]^2\bigr)$ exactly, and
Eq.~\eqref{eq:integer-h} turns this into
\begin{equation}
 |h(y_J)|=B_3:=dw_J-1.
\end{equation}
All three increase with $J$, because the Catalan ratio
$\Cat_{J+1}/\Cat_J=(4J+2)/(J+2)$ outgrows the rational
prefactors and $w_{J+1}/w_J>1$. At $J=6$ they equal
$132/91$, $385/298$ and $301/22$, so
\begin{equation}
 |h(y_j)|\ge\frac{385}{298}>1,\qquad J\ge6,\quad 0\le j\le J,
\end{equation}
and by Eq.~\eqref{eq:integer-h} no lower weight vanishes for
$J\ge6$. For $J\le5$ the bound $B_2$ is below one, and
$|h(y_j)|=1$ indeed occurs at the four exceptional nodes.

\paragraph*{Example.}
For $J=2$, Eq.~\eqref{eq:hdata} gives $h(12)=-1$, $h(20)=1$ and
$[y^2]h=1/48$, hence $h(y)=y^2/48-5y/12+1$ and
$\mathcal R(y)=5+30/(y-12)+30/(y-20)$. The first root is
$\rho_1=10-\sqrt{52}\approx2.79\in(2,3)$. The lower-node values
$h(0)=1$, $h(2)=1/4$, $h(6)=-3/4$ reproduce
$(w_0,w_1,w_2)=(0,1/4,7/20)$; the vanishing $w_0^{(2)}$ is the
case $h=+1$.

\subsection{Half-integer isospin}
\label{sec:half-integer}

Let $J=L+\tfrac12$ with integer $L\ge0$, so $d=2L+2$. The
scalars form a complex multiplet, and the direct and crossed
channels initially carry independent coefficients,
$\mathbb{P}_0^t=\sum_{j\le L}(a_j\mathbb{P}_j^s+b_j\mathbb{P}_j^u)$. The odd-$K$ matching
equations force a polynomial of degree at most $L$ to vanish at
$L+1$ nodes, hence identically, so $a_j=b_j=:w_j$; the even
equations give
\begin{equation}
 \frac2d\sum_{j=0}^L(2j+1)w_jR_j(y)
 =\prod_{k=1}^L\Bigl(1-\frac{y}{y_{2k}}\Bigr).
 \label{eq:half-even-interpolation}
\end{equation}

The crossing identity now carries the opposite sign, and the
vanishing upper weights fix the values of $h$ at $L+1$ points:
\begin{equation}
 dw_j=1+(-1)^jh(y_j),\qquad
 h(y_r)=(-1)^{r+1}\quad(L+1\le r\le2L+1).
\end{equation}
A polynomial of degree $L$ is determined by $L+1$ values, so $h$
is the Lagrange interpolant
\begin{equation}
 h(y)=\sum_{r=L+1}^{2L+1}(-1)^{r+1}\lambda_r(y),\qquad
 \lambda_r(y)=\prod_{\substack{k=L+1\\k\ne r}}^{2L+1}
 \frac{y-y_k}{y_r-y_k}.
\end{equation}
At a lower node every term of this sum has the same sign: each
factor $y_j-y_k$ is negative, and counting the negative factors
in the denominators gives
$\operatorname{sgn}\lambda_r(y_j)=(-1)^{r-L-1}$, which the
alternating prescribed values exactly compensate. Therefore
\begin{equation}
 |h(y_j)|=\sum_{r=L+1}^{2L+1}|\lambda_r(y_j)|>1,\qquad L\ge1,
\end{equation}
the inequality being strict because the Lagrange weights sum to
one while having both signs. Hence $dw_j=1\pm|h(y_j)|\ne0$. For
$L=0$, $h=1$ and $w_0=1$. For $J=\tfrac32$, $h(y)=(y-9)/3$ is the
line through $h(6)=-1$ and $h(12)=1$, and $h(0)=-3$, $h(2)=-7/3$
give $(w_0,w_1)=(-\tfrac12,\tfrac56)$.

Every lower-isospin coefficient is therefore nonzero for every
half-integer $J$, in both physical channels.











\section{Crossing nulls for the exceptional representations}
\label{sec:nulls}

Assume that all massive exchanges have $j\le J$. Define
\begin{equation}
 \langle f\rangle_j^{(r)}
 =\sum_\ell\int_{\Delta^2}^{\infty}
 \frac{f(x_\ell)}{\mu^r}\,d\rho_{j,\ell},
 \qquad x_\ell=\ell(\ell+D-3),
 \label{eq:moment-notation}
\end{equation}
and
\begin{equation}
 \Pi_2^+(x)=x(x-2D+2),\qquad
 \Pi_2^-(x)=(x-D+2)(x-3D),\qquad
 \Pi_3^-(x)=\Pi_2^-(x)(x-5D-10).
 \label{eq:higher-spin-polynomials}
\end{equation}
These weights are nonnegative on even, odd, and odd spins,
respectively. They vanish at spins $(0,2)$, $(1,3)$, and
$(1,3,5)$, and are positive above their last zero.
The main-text moments are
$\mathcal N_j=\langle\Pi_2^+\rangle_j^{(7)}$ for even $j$
and $\mathcal N_j=\langle\Pi_2^-\rangle_j^{(7)}$ for odd $j$.

\paragraph*{$J=3$.}
The coefficient $w_3^{(3)}=3/7$ requires unbounded spin
in $j=3$. The two degree-seven crossing nulls are
\begin{equation}
 3\langle\Pi_2^-\rangle_1^{(8)}
 =14\langle\Pi_2^-\rangle_3^{(8)}>0,
 \label{eq:j3odd-higher}
\end{equation}
and
\begin{align}
 \langle\Pi_2^+(23x-103D-114)\rangle_2^{(8)}
 ={}&5\langle\Pi_3^-\rangle_1^{(8)}+35(D+2)\langle\Pi_2^-\rangle_3^{(8)}>0.
 \label{eq:main-j3-second}
\end{align}
Both right-hand sides are strictly positive because $j=3$
has unbounded spin. The first identity therefore requires
$j=1$ with $\ell\ge5$. In the second, the left-hand weight
vanishes at $\ell=0,2$, is negative at $\ell=4$, and is
positive only for even $\ell\ge6$. Hence $j=2$ must occur
with $\ell\ge6$.

\paragraph*{$J=5$.}
The degree-six crossing null is
\begin{equation}
 4\mathcal N_2=9(\mathcal N_3+\mathcal N_4)>0.
 \label{eq:j5new-higher}
\end{equation}
Here $w_3^{(5)}=13/12$ requires unbounded spin in $j=3$,
so $\mathcal N_3>0$. Thus $j=2$ must occur with $\ell\ge4$.

These three identities exclude the $J=3,5$ omissions.
The remaining $J=2$ singlet omission is excluded by
Eq.~\eqref{eq:j2-degree-six}, even when higher isospins
are present.













\section{The singlet identity and its hierarchy}
\label{sec:singlet}

The identity~\eqref{eq:main-singlet-p-balance} and the zeros of
its spin weights both follow from the residue theorem applied to
rational forms with poles of order $p+1$ at two points exchanged
by crossing.

\subsection{The crossing identity at every order}

For integer $J\ge1$ use a real orthonormal basis of dimension
$d=2J+1$, and let $B=\mathbf A-\mathbf A_{\rm light}$,
Eq.~\eqref{eq:full-light}, viewed as a matrix
$B_{\rm pair}^{ab;cd}=B_{\rm legs}^{abdc}$ from the incoming pair
$(ab)$ to the outgoing pair $(cd)$, with the convention of
Eq.~\eqref{eq:channel-index-dictionary}. Define
\begin{equation}
 S=\sum_{a,c}B_{\rm pair}^{aa;cc},\qquad
 T=\sum_{a,c}B_{\rm pair}^{ac;ac}.
 \label{eq:ST-main}
\end{equation}
$S$ is the $s$-channel singlet projection,
$S=d\langle\Omega|B_{\rm pair}|\Omega\rangle$ with
$\Omega=d^{-1/2}\sum_a|aa\rangle$, and $T$ is the pair-space
trace, i.e.\ the $t$-channel singlet projection: in $t$-channel
components $B=\sum_KB_K\mathbb{P}_K^t$, $T=dB_0$ and
$S=\sum_K(2K+1)(-1)^KB_K$, the sign being the exchange parity of
a pair of isospin $K$. Crossing gives $S(s,t,u)=S(s,u,t)$ and
$T(s,t,u)=S(t,s,u)$.

In the variable $z=s+t/2$, a state of isospin $j$, spin $\ell$
and squared mass $\mu$ produces the direct pole $(a_\mu-z)^{-1}$
at $s=\mu$ and the crossed pole $(a_\mu+z)^{-1}$ at $u=\mu$,
$a_\mu=\mu+t/2$, Eq.~\eqref{eq:pole-pair}. Contracting its
tensor structure as in Eq.~\eqref{eq:ST-main} gives, with
$r_j=2j+1$ and $\epsilon_j=(-1)^j$,
\begin{equation}
 \begin{array}{c|cc}
 & (a_\mu-z)^{-1}&(a_\mu+z)^{-1}\\ \hline
 S&d\delta_{j0}&\epsilon_jr_j\\
 T&r_j&r_j
 \end{array}
 \qquad\text{times }G_\ell^{(D)}(1+2t/\mu)\,d\rho_{j,\ell}.
 \label{eq:singlet-STpoles}
\end{equation}
The direct pole feeds $S$ only for a singlet, while $T$ counts
every state with multiplicity $r_j$; this is what gives the
singlet the opposite sign below.

\paragraph*{The functional.}
Let $[f]_m$ be the coefficient of $z^mt^{3p-m}$ in the
low-energy expansion of $f$, zero out of range; at fixed total
degree $3p$ all terms share the mass denominator $\mu^{3p+1}$
after the dispersive substitution. We need a crossing relation
among these coefficients involving only $m\ge2p$. Restrict to
the homogeneous degree-$3p$ parts, set $t=1$, and write
$F(s)=S(s,1,-1-s)$, $H(s)=T(s,1,-1-s)$, so that
$F=\sum_m[S]_mz^m$ and $H=\sum_m[T]_mz^m$ with $z=s+\frac12$.
Crossing and homogeneity give
\begin{equation}
 H(-1-s)=H(s),\qquad H(s)=s^{3p}F(1/s).
 \label{eq:FH-crossing}
\end{equation}
The reflection $s\mapsto-1-s$ exchanges $s=0$ with $s=-1$, the
inversion $s\mapsto1/s$ exchanges $s=0$ with $s=\infty$; high
coefficients $[f]_m$ live at $z=\infty$, and the residue theorem
transports them to $s=0,-1$, where crossing acts. Introduce
\begin{equation}
 \omega_S=\frac{(s+2)F(s)\,ds}{[s(s+1)]^{p+1}},\qquad
 \omega_T=\frac{(s+\tfrac12)H(s)\,ds}{[s(s+1)]^{p+1}}.
 \label{eq:singlet-residue-forms}
\end{equation}
The poles of order $p+1$ select the coefficients with $m\ge2p$:
with $s(s+1)=z^2-\frac14$,
\[
 [s(s+1)]^{-p-1}=z^{-2p-2}\sum_{i\ge0}c_i\,z^{-2i},\qquad
 c_i=\binom{p+i}{i}4^{-i},
\]
and the numerators are $zH$ and $(z+\frac32)F$, so the
coefficients of $z^{-1}dz$ at infinity are $\sum_ic_i[T]_{2p+2i}$
and $\sum_ic_i([S]_{2p+2i}+\frac32[S]_{2p+2i+1})$. Hence
\begin{equation}
 \begin{split}
 \mathcal L_p[S,T]&:=\sum_{i=0}^{\lfloor p/2\rfloor}
 c_i\left([T]_{2p+2i}-[S]_{2p+2i}-\tfrac32[S]_{2p+2i+1}\right)\\
 &\phantom{:}=\operatorname{Res}_\infty\omega_S
 -\operatorname{Res}_\infty\omega_T,
 \end{split}
 \label{eq:singlet-towerfunctional}
\end{equation}
the sum stopping because $[f]_m=0$ for $m>3p$. The prefactors
$s+\frac12$ and $s+2$ make crossing map one form onto the other:
by Eq.~\eqref{eq:FH-crossing} the reflection leaves $\omega_T$
invariant, so
$\operatorname{Res}_0\omega_T=\operatorname{Res}_{-1}\omega_T$,
and the inversion $s=1/v$ gives $\omega_T=-\frac12\omega_S(v)$, so
$\operatorname{Res}_{s=0}\omega_T=-\frac12\operatorname{Res}_{v=\infty}\omega_S$.
The residue theorem for $\omega_T$ then yields
\begin{equation}
 \operatorname{Res}_\infty\omega_T
 =-2\operatorname{Res}_0\omega_T
 =\operatorname{Res}_\infty\omega_S,
\end{equation}
i.e.\ $\mathcal L_p[S,T]=0$.

\subsection{Spin weights and the singlet conclusion}

By Eqs.~\eqref{eq:singlet-STpoles} and \eqref{eq:central-finite},
the coefficient of $z^mt^n$ in $T$ is
$\sum_{j,\ell}\int r_j[1+(-1)^m]\,Q_{mn}(x_\ell)\,\mu^{-m-n-1}
d\rho_{j,\ell}$, and in $S$ the same with $r_j[1+(-1)^m]$
replaced by $d\delta_{j0}+(-1)^m\epsilon_jr_j$, the sign
$(-1)^m$ coming from the crossed pole. Here
$x_\ell=\ell(\ell+D-3)$, $Q_{mn}$ of Eq.~\eqref{eq:jetkernel} is
the coefficient of $v^n$ in $G_\ell^{(D)}(1+2v)/(1+v/2)^{m+1}$. Since $m+n=3p$ throughout,
$\mathcal L_p=0$ becomes
\begin{equation}
 \begin{gathered}
 \sum_{j,\ell}\int_{\Delta^2}^\infty
 \frac{\mathcal K_{j,p}(x_\ell)}{\mu^{3p+1}}\,d\rho_{j,\ell}=0,\\
 \mathcal K_{j,p}=(2r_j-d\delta_{j0}-\epsilon_jr_j)\,\mathcal U_p
 -\tfrac32(d\delta_{j0}-\epsilon_jr_j)\,\mathcal V_p,
 \end{gathered}
 \label{eq:singlet-kernel-UV}
\end{equation}
with the even- and odd-order sums
\begin{align}
 \mathcal U_p&=\sum_i c_iQ_{2p+2i,p-2i}
 =[v^p]\frac{(1+v/2)\,G_\ell^{(D)}(1+2v)}{(1+v)^{p+1}},\nonumber\\
 \mathcal V_p&=\sum_i c_iQ_{2p+2i+1,p-2i-1}
 =[v^{p-1}]\frac{G_\ell^{(D)}(1+2v)}{(1+v)^{p+1}},
 \label{eq:UV-closed}
\end{align}
where $Q_{mn}=0$ for $n<0$, $[v^n]$ extracts a power-series
coefficient, and the closed forms follow from
$\sum_ic_i\,[v^2/(1+v/2)^2]^i=(1+v/2)^{2p+2}/(1+v)^{p+1}$.
For even $j$, $\mathcal K_{j,p}=(r_j-d\delta_{j0})K_0$ with
$K_0=\mathcal U_p+\frac32\mathcal V_p$; for odd $j$,
$\mathcal K_{j,p}=3r_jK_1$ with $K_1=\mathcal U_p-\frac12\mathcal V_p$.
Only the singlet coefficient, $r_0-d=-2J$, is negative.

The spin weights are residues of the same type as
Eq.~\eqref{eq:singlet-residue-forms},
\begin{equation}
 K_\epsilon(x_\ell)=\operatorname{Res}_{v=0}
 \frac{(1+2v)^{1-\epsilon}G_\ell^{(D)}(1+2v)\,dv}
      {[v(v+1)]^{p+1}},\qquad \epsilon=\ell\bmod2.
 \label{eq:singlet-root-residue}
\end{equation}
Under $v\mapsto-1-v$ the factor $1+2v$ changes sign and $G_\ell$
acquires $(-1)^\ell$, so the form is invariant and its residues
at $0$ and $-1$ are equal. For $\ell<2p$ the numerator has
degree at most $2p-1$ against $2p+2$ in the denominator, so the
residue at infinity vanishes and both finite residues are zero:
$K_\epsilon(x_\ell)=0$ for
$\ell=\epsilon,\epsilon+2,\dots,\epsilon+2p-2$. Since
$K_\epsilon$ has degree at most $p$ in $x_\ell$, these $p$ roots
and the leading coefficient of Eq.~\eqref{eq:angular-series} fix
it:
\begin{equation}
 K_\epsilon(x)=\frac{\Pi_{p,\epsilon}(x)}{p!\,((D-2)/2)_p},
 \qquad
 \Pi_{p,\epsilon}(x)=\prod_{r=0}^{p-1}[x-x_{\epsilon+2r}].
 \label{eq:K-closed}
\end{equation}
On the allowed spins, $\Pi_{p,j\bmod2}(x_\ell)$ vanishes at the
first $p$ spins and is positive above them. Altogether,
\begin{equation}
 \mathcal K_{j,p}(x)=
 \frac{(2j+1)(2-(-1)^j)-d\delta_{j0}}{p!\,((D-2)/2)_p}\,
 \Pi_{p,j\bmod2}(x),
 \label{eq:singlet-tower-kernel}
\end{equation}
with mass denominator $\mu^{3p+1}$; dropping the common positive
factor gives Eq.~\eqref{eq:main-singlet-p-balance}. At $p=2$ and
$D=4$, $\Pi_{2,j\bmod2}=\pi_j$ and
$\mathcal N_j^{(2)}=\mathcal N_j$, recovering
Eq.~\eqref{eq:j2-degree-six}, including the unrestricted $J=2$
identity.

\subsection{Any compact group, including complex matter}
\label{sec:universal-singlet}

The construction used only the invariant pair $\Omega$ and the
exchange sign $(-1)^\ell$ of a state of spin $\ell$. For a real
scalar representation $V$ of dimension $n_V$ of a compact group
$G$, let $\mathsf F$ interchange the factors of $V\otimes V$ and
define
\begin{equation}
 |\Omega\rangle=\frac1{\sqrt{n_V}}\sum_a|aa\rangle,\qquad
 \mathbb{P}_\Omega=|\Omega\rangle\langle\Omega|,\qquad
 \mathbb{P}_+=\frac{\Id+\mathsf F}{2}-\mathbb{P}_\Omega,\qquad
 \mathbb{P}_-=\frac{\Id-\mathsf F}{2}.
 \label{eq:universal-pair-projectors}
\end{equation}
A compact group admits an invariant positive metric, so its
action is orthogonal in this basis and $\Omega$ is invariant.
These positive projectors commute with $G$ and resolve the pair
space into the singlet $\Omega$, the other symmetric states, and
the antisymmetric states, even if $V$ is reducible or contains
other invariant pairs.

With $S=n_V\langle\Omega|B_{\rm pair}|\Omega\rangle$ and
$T=\Tr B_{\rm pair}$, the crossing relations and $\mathcal L_p$
are unchanged. At a pole with residue operator $R$, set
$\tau=\Tr R$ and $\sigma=\langle\Omega|R|\Omega\rangle$. The
direct and crossed $S$ weights are $n_V\sigma$ and
$\Tr(R\mathsf F)=(-1)^\ell\tau$, both $T$ weights are $\tau$.
Repeating the computation of the kernel, with
$\tau=\sigma+\Tr(\mathbb{P}_+R)$ for even $\ell$ and $\sigma=0$,
$\tau=\Tr(\mathbb{P}_-R)$ for odd $\ell$, gives
\begin{equation}
 \boxed{(n_V-1)\,\mathcal I_{\Omega,p}
       =\mathcal I_{+,p}+3\,\mathcal I_{-,p},}
 \label{eq:general-singlet-null}
\end{equation}
\begin{equation}
 \mathcal I_{a,p}=\sum_\ell\int_{\Delta^2}^{\infty}
 \frac{\Pi_{p,\epsilon_a}(x_\ell)}{\mu^{3p+1}}\,d\sigma_{a,\ell},
 \qquad d\sigma_{a,\ell}=\Tr(\mathbb{P}_a\,d\Gamma_\ell)\ge0,
 \qquad \epsilon_\Omega=\epsilon_+=0,\quad\epsilon_-=1,
 \label{eq:universal-positive-moments}
\end{equation}
with $d\Gamma_\ell$ the positive residue measure. For $n_V>1$,
absence of high-spin $\Omega$ residues would force all three
nonnegative moments to vanish, hence every coupled spin to lie
below $2p$, contradicting the graviton relation as in the main
text. For $n_V=1$ every exchange is a singlet and the
bounded-spin contradiction suffices.

Factorization gives $R_X=g_X^\dagger g_X$ with
$g_X:V\otimes V\to W_X$ commuting with $G$; since $\Omega$ is
invariant, so is $g_X\Omega$, and $\sigma=\|g_X\Omega\|^2$
receives contributions only from singlet components of $W_X$.
Complex or pseudoreal matter is included by writing
$\phi=(X+iY)/\sqrt2$ and keeping the physical antiparticles: a
unitary representation becomes orthogonal on its real
components, and for complex dimension $d_R$ the invariant pair is
\begin{equation}
 |\Omega\rangle=\frac1{\sqrt{2d_R}}\sum_i
 (|i\bar i\rangle+|\bar i i\rangle).
 \label{eq:scalar-realification}
\end{equation}
In this real basis the singlets couple in the neutral
$R\otimes\overline R$ channel of the original multiplet. The
theorem fixes neither every even spin nor a mass--spin
trajectory, and need not identify the towers seen by different
scalar probes.







\section{The two-channel obstruction}
\label{app:explicit-dual}

Within a class of amplitudes built from Euler integrals, we show
that massive exchanges confined to $j\le L$ force every building
block to obey the cutoff in both of its channels,
Eq.~\eqref{eq:Euler-color-intersections}, and that for $L=J\ge2$
this leaves no room for the singlet graviton pole.

\subsection{The cutoff in both channels}
\label{app:Euler-slopes}

In units where the common slope is one, so that all poles lie at
nonnegative integers, write
\begin{equation}
 \mathcal A=\Phi_{st}+\Phi_{su}+\Phi_{tu}+\mathcal C,
 \label{eq:Euler-channel-decomposition}
\end{equation}
with $\mathcal C$ pole-free and
\begin{equation}
 \Phi_{st}(s,t)=
 \int_0^1dx\,x^{-t-1}(1-x)^{-s-1}
 \sum_{h+k\le N}t^hs^k f_{hk}^{st}(x),
 \label{eq:regular-Euler-class}
\end{equation}
where $N$ is finite and the invariant-tensor functions
$f_{hk}^{st}$ are analytic. The class contains the
amplitude~\eqref{eq:explicit-J1-amplitude}, with $N=2$ and
constant $f_{hk}$, and any finite sum of polynomially dressed
beta functions with nonnegative integer shifts of their
arguments. We assume that every massive pole in channel $c$ has
its residue in $\mathcal L_c=\operatorname{span}\{\mathbb{P}_j^c:0\le j\le L\}$.

\paragraph*{Residues as Taylor coefficients.}
The $t$-channel poles of $\Phi_{st}$ come from the endpoint
$x=0$: a term $c\,x^n$ of the rest of the integrand gives
$\int_0^\epsilon dx\,x^{n-t-1}=\epsilon^{n-t}/(n-t)$, a simple
pole at $t=n$. To absorb the explicit factors $t^h$, which equal
$n^h$ at this pole, integrate by parts using
$t\,x^{-t-1}=-\partial_x(x^{-t})$: each factor $t$ becomes the
operator $x\partial_x$ on the remaining integrand, which
reproduces $n$ on $x^n$. In the $tu$ block first substitute
$x\to x/(1+x)$ and $u=-s-t$, which moves all $t$ dependence into
$x^{-t-1}$. The result is
\begin{align}
 \Phi_{st}(s,t)
 &=\int_0^1dx\,x^{-t-1}(1-x)^{-s}\,U(x,s),
 \nonumber\\
 \Phi_{tu}(t,-s-t)
 &=\int_0^\infty dx\,x^{-t-1}(1+x)^{-s}\,V(x,s),
 \label{eq:Euler-transformed-blocks}
\end{align}
with $U,V$ polynomials in $s$ whose coefficients are analytic
near $x=0$ and along the open integration interval. Hence the
coefficient of $1/(n-t)$ in $\Phi_{st}+\Phi_{tu}$ is
\begin{equation}
 c_n(s)=[x^n]\bigl\{(1-x)^{-s}U(x,s)+(1+x)^{-s}V(x,s)\bigr\},
 \label{eq:Euler-slope-residue}
\end{equation}
a polynomial in $s$. The $su$ block and $\mathcal C$ have no
$t$-channel poles for generic $s$.

\paragraph*{Separating the two blocks.}
Fix a forbidden isospin $j>L$ and let the subscript $j$ denote
the $\mathbb{P}_j^t$ component. By hypothesis $c_{n,j}(s)=0$ for all
$n\ge1$ and all $s$, so the generating function of the residues
collapses to its $n=0$ term,
\begin{equation}
 (1-z)^{-s}U_j(z,s)+(1+z)^{-s}V_j(z,s)
 =\sum_{n\ge0}c_{n,j}(s)z^n=c_{0,j}(s),
 \label{eq:Euler-massless-remainder}
\end{equation}
for all $s$ and all sufficiently small $z>0$, where $c_{0,j}$
would be the coefficient of a massless pole $1/(-t)$. At fixed
$0<z<1$ the three terms are exponentials $e^{\lambda s}$ with
the distinct rates $\lambda=-\log(1-z)>0$, $-\log(1+z)<0$ and
$0$, multiplied by polynomials in $s$. Exponentials with distinct
real rates are independent over polynomials.
Hence
\begin{equation}
 U_j(z,s)=V_j(z,s)=0,\qquad c_{0,j}(s)=0.
 \label{eq:Euler-block-vanishing}
\end{equation}

\paragraph*{Conclusion.}
$U_j$ vanishes near $x=0$, hence by analyticity on the whole
integration interval, and likewise $V_j$; the exact
representation~\eqref{eq:Euler-transformed-blocks} then gives
$\Phi_{st,j}=\Phi_{tu,j}=0$ identically, pole-free parts
included. Repeating for the $s$-channel poles (endpoint $x=1$ of
$\Phi_{st}$, and $\Phi_{su}$) and for the $u$-channel poles
($\Phi_{su}$ and $\Phi_{tu}$),
\begin{equation}
 \Phi_{st}\in\mathcal L_s\cap\mathcal L_t,\quad
 \Phi_{su}\in\mathcal L_s\cap\mathcal L_u,\quad
 \Phi_{tu}\in\mathcal L_t\cap\mathcal L_u.
 \label{eq:Euler-color-intersections}
\end{equation}
Only $\Phi_{st}$ and $\Phi_{tu}$ have a pole at $t=0$, whose
residue is the pure singlet $g_Gs^2\mathbb{P}_0^t$ at order $s^2$. Each
block contributes to this coefficient within its own subspace,
so
\begin{equation}
 \mathbb{P}_0^t\in
 (\mathcal L_t\cap\mathcal L_s)
 +(\mathcal L_t\cap\mathcal L_u).
 \label{eq:Euler-graviton-membership}
\end{equation}

\subsection{The obstruction for $L=J$}
\label{app:Euler-LJ}

By Eq.~\eqref{eq:main-sixj-conversion},
$X=\sum_{j=0}^Jc_j\mathbb{P}_j^s\in\mathcal L_s$ has $t$-components
$X_K=(-1)^Kp_X(y_K)/d$, where $p_X=\sum_j(2j+1)c_jR_j$ is an
arbitrary polynomial of degree at most $J$. It lies in
$\mathcal L_t$ iff $p_X$ vanishes at $y_{J+1},\dots,y_{2J}$,
i.e.\ iff $p_X\propto Q(y)=\prod_{r=J+1}^{2J}(1-y/y_r)$, the
polynomial of Appendix~\ref{sec:allJ}. Hence
$\mathcal L_s\cap\mathcal L_t$ and $\mathcal L_t\cap\mathcal L_u$
are one-dimensional, spanned by
\begin{equation}
 T_{st}=\sum_{K=0}^J(-1)^KQ(y_K)\,\mathbb{P}_K^t,\qquad
 T_{tu}=\sum_{K=0}^JQ(y_K)\,\mathbb{P}_K^t,
 \label{eq:Euler-Tst}
\end{equation}
the sign being absent for $u$ because
$(M_{u,t})_{jK}=(-1)^K(M_{s,t})_{jK}$. For $J=1$ these are
$3\mathbb{P}_0^t\mp2\mathbb{P}_1^t$ up to normalization, the tensors of
Eq.~\eqref{eq:explicit-J1-amplitude}. Writing
$\mathbb{P}_0^t=\alpha T_{st}+\beta T_{tu}$ and comparing $\mathbb{P}_K^t$
components gives $\alpha+\beta=1$ at $K=0$ and
$[(-1)^K\alpha+\beta]\,Q(y_K)=0$ for $1\le K\le J$, where
$Q(y_K)\ne0$. For $J=1$ this yields $\alpha=\beta=\tfrac12$, the
solution~\eqref{eq:integer-example-t-inverse}. For $J\ge2$ the
conditions at $K=1,2$ give $\alpha=\beta=0$, a contradiction.
Within the
class~\eqref{eq:Euler-channel-decomposition}--\eqref{eq:regular-Euler-class},
a graviton pole therefore requires a massive exchange with $j>J$
for every $J\ge2$, whatever the coefficient functions;
Eq.~\eqref{eq:explicit-J2-first-t-residue} is the $J=2$ instance.

\end{document}